\documentstyle[manuscript,aps]{revtex}
\begin{document}

\title{On Evolution Laws Taking Pure States to Mixed States in Quantum
Field Theory}
\author{William G. Unruh\\
         {\it CIAR Cosmology Program }\\
         {\it Department of Physics}\\
         {\it University of British Columbia}\\
         {\it Vancouver, BC, Canada V6T 2A6} \\
{\ }\\
Robert M. Wald\\
         {\it Enrico Fermi Institute and Department of Physics}\\
         {\it University of Chicago}\\
         {\it 5640 S. Ellis Avenue}\\
         {\it Chicago, Illinois 60637-1433}}
\date{}

\maketitle
\begin{abstract}
It has been argued that any evolution law taking pure states to mixed states
in quantum field theory necessarily gives rise to
violations of either causality or energy-momentum conservation,
in such a way as to have unacceptable consequences for ordinary
laboratory physics. We show
here that this is not the case by giving a simple class of examples of
Markovian
evolution laws where rapid evolution from pure states to mixed states
occurs for a wide class of states with appropriate properties at the
``Planck scale", suitable locality and causality properties
hold for all states, and the deviations from ordinary
dynamics (and, in particular,
violations of energy-momentum conservation) are
unobservably small for all
states which one could expect to produce in a laboratory. In addition, we
argue (via consideration of other,
non-Markovian  models) that conservation of energy and
momentum for all states
is not fundamentally incompatible with causality in dynamical
models in which pure states evolve to mixed states.

\end{abstract}
\newpage

	One of the most striking ramifications of the discovery that
black holes should radiate as black bodies \cite{haw} is the implication
that a black hole should completely evaporate within a finite time, and
that, in this process, an initially pure quantum state
should evolve to a mixed state (see, e.g., \cite{wald} for a review of
arguments leading to this conclusion). This prediction of a ``loss of
quantum coherence" is derived in the semiclassical approximation
by applying the ordinary dynamical evolution laws locally
to the quantum field, so no violation of any of the principles of local
quantum field theory occurs in this context. Indeed, in the semiclassical
approximation, the loss of quantum coherence is directly attributable to
the failure of the final time surface in the spacetime representing the
evaporating black hole to
be a Cauchy surface; exactly the same sort of phenomenon
occurs when one considers the evolution of a free, massless field in Minkowski
spacetime with the initial surface chosen as a ordinary hyperplane, but
the final surface chosen as a hyperboloid.

Nevertheless, it seems natural to expect that when one goes beyond
the semiclassical approximation, the possibility of
loss of quantum coherence in black hole formation
and evaporation should give rise to a significant (in principle)
modification of the ordinary, local, dynamical
evolution laws: For almost any
initial quantum state, one would expect there to be a nonvanishing
amplitude for black hole formation and evaporation to occur -- at
at least at a highly
microscopic (e.g., Planckian) scale -- thereby giving rise to a nonvanishing
probability for evolution from pure states to mixed states
\cite{EHNS}, \cite{huet-peskin}.
 One would expect the deviations from the
ordinary dynamical evolution laws to be negligibly small for all states
normally accessible to laboratory experiments, but it would seem
reasonable to expect large deviations to occur when, for example, the state
of the quantum field is such that there is a
substantial probability to produce black holes at the Planck scale.

It should be emphasized that any such modification of the ordinary
dynamical laws which permits pure states to evolve to mixed states
should fundamentally involve quantum gravity and, hence, a priori,
there is no reason to expect it to be possible to
adequately describe such effects by some ``effective theory" in
which spacetime structure is treated classically. A good analogy to bear
in mind in this regard is the classical theory of electromagnetism with
point particles. Here, when radiation reaction effects are included,
the theory is found either to violate causality or to have unacceptable
consequences for laboratory physics (namely, run-away solutions).
In this case, one does not normally view this nonexistence of an
acceptable ``effective classical theory" as an indication that quantum
electrodynamics must be plagued by similar pathologies; rather, the
usual interpretation of this situation is simply that a purely classical
treatment of point particles is inadequate to describing radiation
reaction phenomena. Similarly, a purely classical treatment of spacetime
structure may be inadequate to describe phenomena in which pure
states evolve to mixed states.

Indeed, it should be noted that black hole evaporation is an unusual
process, in that during the formation
of the black hole, the energy is transfered from the matter to the
gravitational
field of the black hole. The entropy of the matter, however, remains with
the matter and is lost down the black hole, taking no energy with it. By
contrast, for ordinary systems, the
interactions with the environment which produce decoherence
normally exchange both energy and entropy. Thus, the loss of coherence
of a system normally is accompanied
by energy non-conservation. What is needed in order to
properly model what is believed to occur in the black
hole case is to have an environment
which can increase the entropy of a system, while at the same time
exchanging energy with the system during only a limited interval of time,
such that no net exchange of energy occurs. Thus, it should not be
surprising that classical spacetime models which have not been carefully
designed to do this (see the Appendix) will face difficulties
with energy conservation. However, this does not imply that similar
difficulties with energy conservation need occur in processes involving
black holes.

In any case, it is of interest to know if
-- in the context of theories where spacetime
structure is treated classically --
there exists any difficulty, in principle, in finding mathematically well
defined ``effective dynamical evolution laws"
such that a suitable class of
pure states can rapidly evolve to mixed states, but no inconsistencies
with known laboratory physics occur. If a fundamental difficulty is
present, then -- despite the comments in the
previous two paragraphs -- this could conceivably indicate the presence
of a similar difficulty in quantum gravity.
This issue was addressed by Banks, Susskind, and Peskin \cite{bsp}, who,
following an earlier analysis of Ellis et al \cite{EHNS},
argued that a serious difficulty of principle does exist. They concluded that
any dynamical evolution law which takes pure states to mixed states
(with appropriately large probability for suitable states) must give rise to
unacceptably large violations of causality or energy-momentum
conservation at the scales of laboratory physics (however see \cite{EMN}).
Although these authors
did not claim to provide a complete proof of this conclusion, their arguments
have gained widespread acceptance and appear to underlie many efforts
to modify the picture of black hole formation and evaporation provided
by the semiclassical approximation, so that an initially pure quantum
state will remain pure in that process.

In this paper, we shall re-examine the arguments of \cite{bsp} and draw
the opposite conclusion: We will consider what is, in essence, simply a
subclass of the models considered in \cite{bsp} with good causal
properties, and will show that they
can be adjusted to yield an arbitrarily rapid loss of quantum coherence for
states with suitable properties at, say, the Planck scale,
but produce a negligible
deviation from ordinary dynamical evolution for states which can be
produced in
laboratories. Thus, although in these models violations of energy and
momentum conservation presumably would occur
(as argued in \cite{bsp}) and Lorentz invariance presumably also
would fail \cite{S}, we shall
show that (contrary to the claims made in \cite{bsp})
there is no difficulty confining such pathologies to the
``Planckian states", which are not accessible to ordinary laboratory physics.

Although we believe that the models considered in the body of the paper
below would predict violations of energy-momentum conservation if one
did scattering experiments with particles of Planck scale energy,
it should be emphasized that these models
(as well as the somewhat more general models considered in \cite{bsp})
encompass only "Markovian" models, where the equation of motion
governing the time evolution of states is local in time; more precisely,
the time evolution map has the structure of a dynamical semigroup
(see, e.g., \cite{L}). Since a black hole should have a long timescale
``memory" (stored in its external
gravitational field) of the amount of energy that went into it, one
would not expect an effective evolution law modeling the process of black
hole formation and evaporation to be Markovian in nature.
In the Appendix, we shall consider some alternative,
non-Markovian models. Although these models are not satisfactory
as models of the black hole formation and evaporation process, they
serve the purpose of showing that causality and
energy-momentum conservation are not fundamentally in conflict for
evolution laws taking pure states to mixed states,
i.e, one can construct classical spacetime models
in which pure states evolve to mixed states and energy-momentum
conservation and causality hold on all scales, not merely on the scales
of laboratory physics.

To construct our Markovian
models displaying rapid loss of quantum coherence for
``Planckian states", but negligible
deviation from ordinary dynamics for ``laboratory states",
we start with an ordinary, causal, unitary,
local quantum field theory in Minkowski spacetime -- such
as a free Klein-Gordon field -- with dynamics determined by
a Hamiltonian $H$. We then consider the following class of modified
dynamical laws,
\begin{equation}
\dot{\rho} = -i[H,\rho] - \sum_i \lambda_i (Q_i \rho +
\rho Q_i - 2Q_i \rho Q_i)
\label{eom}
\end{equation}
where each $\lambda_i$ is positive and
where each $Q_i$ is an orthogonal projection operator (i.e., $Q_i^\dagger =
Q_i$ and $Q_i^2 = Q_i$).
This modified dynamical law is simply a specialization of the general
form of a Markovian evolution law given in eq. (4.3) of \cite{L} to the case
where the operators $V_j$ appearing in that equation are projection
operators; it also corresponds to eq.(9) of \cite{bsp}, specialized to the
case where their matrix $h_{\alpha\beta}$ is diagonal and their $Q's$
are projection operators.
We do not impose any additional conditions upon
the different $Q_i 's$ at this stage, although we shall impose an
additional locality restriction below to assure suitable
causality properties of the theory; in particular, note that the
$Q_i's$ are not assumed to commute.

Taking the trace of eq.(\ref{eom}), we
immediately obtain $tr \dot{\rho} = 0$. Furthermore, the arguments of
\cite{bsp} show that
eq.(\ref{eom}) preserves the positivity of $\rho$, and that it decreases the
purity of states in the sense that
$d/dt [tr(-\rho \ln \rho)] \geq 0$. Thus, eq.(\ref{eom}) evolves pure
states into mixed states in such a way as to conserve probability and keep
all probabilities positive.

Since eq.(\ref{eom}) is linear in $\rho$, it defines a linear time evolution
operator $\$(t_0,t)$ on density matrices, so that, in index notation,
we have
\begin{equation}
{\rho^A}_B (t) = \${^A}{_B}{_C}{^D}(t_0,t) {\rho^C}_D (t_0)
\label{evol}
\end{equation}
The Heisenberg representation version of this dynamics (where the states
are viewed as fixed in time) is obtained by evolving
each observable, $A$, by a suitable
transpose of $\$$, namely
\begin{equation}
{A^A}_B (t) = \${^C}{_D}{_B}{^A}(t_0,t) {A^D}_C (t_0)
\label{hevol}
\end{equation}
This corresponds to the following Heisenberg equation of motion for
$A$,
\begin{eqnarray}
\dot{A} & = & +i[H, A] - \sum_i \lambda_i (Q_i A +
A Q_i - 2Q_i A Q_i) \nonumber \\
 & = & +i[H, A] + \sum_i \lambda_i [Q_i, [A, Q_i]]
\label{heom}
\end{eqnarray}
Equation (\ref{heom}) will be useful for our analysis below of the locality
properties of the model \cite{com1}.

Some insight into the nature of the dynamics defined by
eq.(\ref{eom}) can be gained by considering the special case where
$H = 0$ and only one of the $Q_i's$ (denoted $Q$)
is present. In this case, we decompose
$\rho$ into a $2 \times 2$ block matrix with respect to the subspace
defined by $Q$ and its orthogonal complement. It then is easily seen
that the evolution law
eq.(\ref{eom}) leaves the diagonal blocks of $\rho$ unchanged but causes
the off-diagonal blocks to exponentially decay away. Thus, in this
case, eq.(\ref{eom}) corresponds simply
to a decoherence between the subspace
associated with $Q$ and its orthogonal complement.
Of course, a much richer
dynamics can occur when $H \neq 0$ and when
many noncommmuting $Q_i's$ are present.

We now shall further specialize eq.(\ref{eom})
to give the model suitable locality properties.
This will allow us to keep violations of causality under control (even
for ``exotic" states where rapid evolution from pure to mixes states
occurs) and enable us to ensure that possible exotic phenomena occurring
in distant regions of the universe will not affect observations in our
laboratories.
Let ${\cal R}$ be any region of space and
let $R$ be any local field observable for this region. (A good example of
such an $R$ of possible relevance to black hole formation issues would be
obtained by integrating the 4-momentum density of the field over
${\cal R}$ and then squaring it.)
Let $Q$ be a projection operator
onto a subspace spanned by eigenvectors of $R$; for definiteness, we
choose a real number, $\alpha$, and take $Q$ to be the projection
operator onto the subspace of eigenvectors of $R$ with eigenvalues
greater than $\alpha$. Now, let
${\cal T}$ be any spatial region disjoint from ${\cal R}$, and let $T$
be any local field observable for ${\cal T}$.
Then $T$ commutes with $R$ and, hence,
$T$ commutes with all of the projection
operators occurring in the spectral resolution of $R$. Hence, in particular,
$[Q,T] = 0$. Since the unmodified, unitary quantum field theory
has causal propagation, the Heisenberg representative of $T$ in the
unmodified theory will commute with $Q$ until such time as a light signal
from ${\cal T}$ can reach
${\cal R}$. Now, consider the modified,
non-unitary Schrodinger dynamics defined by
\begin{equation}
\dot{\rho} = -i[H,\rho] -  \lambda(Q \rho + \rho Q - 2Q \rho Q)
\label{eom'}
\end{equation}
which corresponds to the Heisenberg dynamics
\begin{equation}
\dot{T} = +i[H,T] + \lambda [Q, [T,Q]]
\label{heom'}
\end{equation}
for $T$.
By inspection, the solution to the Heisenberg equation of motion for the
unmodified theory solves eq.(\ref{heom'}) until a light signal
from ${\cal T}$ can reach
${\cal R}$.. This implies that starting from any initial
(globally defined) state at $t = 0$, an observer
in the theory defined by eq.(\ref{eom'}) who makes local measurements
in the region ${\cal T}$ will
not be able to detect any difference from ordinary dynamics until effects
from region ${\cal R}$ can causally propagate to him. In other words, the
dynamics defined by eq.(\ref{eom'}) is entirely ``ordinary"
(and, in particular, local and causal) outside of the
region ${\cal R}$. If we now choose ${\cal R}$ to be so small (e.g., Planck
dimensions) that it is inaccessible to laboratory measurements, no
departures whatsoever from causality and locality will be detectable.

We now restrict our model so that
each projection operator, $Q_i$, appearing in
eq.(\ref{eom}) is a projection operator for a local observable $R_i$
associated with an inaccessibly small region of space ${\cal R}_i$.
(We also
may allow ``large" regions, ${\cal R}_i$, provided that we then choose
the corresponding
$\lambda_i$ to be sufficiently small.) If the different
regions, ${\cal R}_i$, are
allowed to overlap, observable violations of causality could still occur in
the theory (due to ``chain reactions").
However, it is clear that even these potential
violations can be kept under good control by imposing only
mild restrictions on both the overlap of the regions
and the values of the $\lambda_i's$. Thus, we claim that
there is no difficulty in
adjusting the model so that the resulting dynamics will
be observably local and
causal for essentally all states, including those for which the rate of loss
of quantum coherence is large.

The restriction we have placed upon the $Q_i's$, corresponds, in essence,
to a discretized version of
eq.(20) of \cite{bsp}, with the support of their spatial smearing function
$h(x-y)$ chosen to be small. This restriction on the
dynamics also was imposed in
\cite{bsp} to ensure good causal propagation properties of the model.
Hence, we have not, thus far, diverged in any substantial way from the
analysis of \cite{bsp}. The authors of \cite{bsp} then argue further that
violations of energy-momentum conservation must occur. We see
no reason to question this conclusion for the above model, although
in the Appendix we will show  that other,
non-Markovian models need not
suffer from this problem. The authors of \cite{bsp} then
claim that these violations will be so large as to have a drastic effect
upon ordinary laboratory physics. No general arguments are given
in \cite{bsp} to support this assertion, but an example is worked out
which illustrates this phenomenon. However, we now shall
show that there is no difficulty in tailoring the dynamics defined by
eq.(\ref{eom}) (with our above restriction on the $Q_i's$) so that all
violations of energy-momentum conservation (as
well as any other deviations from ordinary dynamics)
would not be detectable in laboratory experiments.

To do so, we identify a subspace, ${\cal H}_L$,
of the Hilbert space of states
corresponding to those states accessible to laboratory physics, and we also
choose a collection of
``inaccessibly small" regions of space ${\cal R}_i$ along with
corresponding observables, $R_i$. The details
of these choices are not of great importance, provided only that (i) the
subspace ${\cal H}_L$ is mapped into itself by the ordinary, unitary
dynamics defined by $H$ and (ii) the regions ${\cal R}_i$ are chosen
sufficiently small that the states in ${\cal H}_L$ restricted to any
${\cal R}_i$ do not differ grossly from some
representative state, say,  the vacuum state $|0>$.
A concrete example of such choices in Klein-Gordon theory would be
to take ${\cal H}_L$ to be the
subspace of states spanned by particles whose mode functions contain
no frequencies higher than $\omega_0$ (where $\omega_0$ is
chosen to be much
higher than is achievable in laboratories) and to take each ${\cal R}_i$ to
have size much smaller than $\omega_0^{-1}$. We now choose (some of)
the $\lambda_i's$ to be as large as we like, thus ensuring that arbitrarily
rapid loss of quantum coherence occurs for some states. Finally, we choose
each $Q_i$ so that each $\epsilon_i \equiv \lambda_i ||Q_i |0>||^2$ is as
small as we wish. This latter condition can easily be achieved by
choosing $\alpha_i$ sufficiently large (so that $Q_i$ is a projection
onto a subspace of extremely large eigenvalues of $R_i$). It is worth
noting that since for typical local observables
$||Q_i |0>||^2$ will fall-off very rapidly with $\alpha_i$ -- for example,
$||Q_i |0>||^2$ will be essentially Gaussian in $\alpha_i$ if $R_i$ is the
field operator smeared over ${\cal R}_i$ -- it normally will not be necessary
to choose $\alpha_i$ to be significantly larger than the scales associated
with ${\cal R}_i$ in order to obtain the desired smallness of $\epsilon_i$.
Since the ``laboratory states" do not differ greatly from $|0>$ with respect
to the observables $R_i$, this will ensure that $\lambda_i tr(\rho Q_i)$
is similarly small for any density matrix
$\rho$ constructed from the
``laboratory states". It then follows immediately that the probability that
eq.(\ref{eom}) will result in an observable difference from the unmodified,
unitary dynamics for ``laboratory states" will be negligible.

It is useful to compare the analysis of the above paragraph to the
illustrative model given in section 5 of \cite{bsp}. In essence, the model
of \cite{bsp}
differs from the above models only in that our projection operator
$Q_i$ is replaced by the squared field operator at a point
(made finite by the imposition of a momentum cutoff at the Planck
scale). However, this squared field operator applied to the vacuum state
yields a state with large norm, so one must choose the parameter $a$
of their model
(corresponding to our $\lambda_i$) to be exceedingly small in order
to avoid affecting ordinary laboratory dynamics. With $a$ chosen to be
this small, non-unitary dynamics
occurs only for states with exotic properties at energy scales far in
excess of the Planck scale. However, this difficulty
of the model could be avoided
by replacing the squared field operator by a suitable
function of the field operator, where the function is chosen
so that it is (significantly) non-zero only for values of the field
which are sufficiently large that vacuum fluctuations to that value of
the field are highly improbably. A step function (yielding a projection
operator) is ideal for this purpose.

In summary, we have shown that even the simple class of
Markovian models
considered here is sufficient to
encompass dynamics which differs
imperceptibly from ordinary dynamics for
all states that have properties similar to the vacuum state at extremely
small scales, but is highly non-unitary
for states that differ greatly from the vacuum state on these scales.
In essence, our analysis differs from \cite{bsp} only in that we have
developed their basic model
sufficiently that one can explicitly see how to
independently control the values of the
quantities $\lambda_i$ (which determine
the maximum rate of loss of quantum coherence) and $\epsilon_i$ (which
determine the probabilities for observing a loss of quantum coherence
for laboratory states). We conclude that there appears to
be no difficulty of principle
in constructing theories which capture the essential features that
might be expected if black hole formation and
evaporation at a highly microscopic scale occurs in the manner suggested
by the semiclassical picture.

We wish to thank Eanna Flanagan for reading the manuscript and making
a number of helpful remarks. This research was supported in
part by the Canadian Institute for
Advanced Research, by the Natural Science and Engineering Research
Council of Canada, and by National Science Foundation
grant PHY-9220644 to the University of Chicago.

{\large \bf Appendix}

The models considered above are Markovian in nature. Such models
arise naturally as an effective dynamics of a system coupled to another
large system (i.e., ``heat bath") in the limit where the relaxation time of
the heat bath goes to zero. The Markovian character of an effective
evolution law makes it difficult to lose coherence while conserving energy
exactly. However, one would not expect the effective dynamics
corresponding to the process of black hole formation and evaporation to
be Markovian, since the black hole should ``remember" (via its external
gravitational field) the amount of energy which was dumped into it,
and it should be able to return this energy via particle creation at
very late times. Indeed, this is especially true if correlations {\em are}
restored during the late stages of black hole evaporation, since this would
require an exceedingly long ``relaxation time" of the black hole system.
In this Appendix, we will analyze two
simple non-Markovian
models for the loss of coherence,
which provide good illustrations of our claim that there need not
be any conflict between loss of coherence,
causality, and energy-momentum conservation. The basic idea of both
models is to have a ``hidden system"
interacting with the given system.
This hidden system will have no energy of its own and therefore
will not be available as either a net source or a sink of energy.
However, the state of that hidden system will affect the behaviour of the
system of interest, in such a way as to produce a loss of quantum
coherence in the system of interest.

For the reasons indicated at the end of this Appendix, the models we treat
here are not satisfactory as models of the black hole evaporation
process. However, it should be noted that the loss of coherence
without energy loss has also been studied in realistic condensed matter
systems\cite{SP}. There the hidden sector which causes the decoherence is
taken to be the nuclear spins of the atoms making up the system. Because
of the weak interactions of the nuclear spins with each other, they
comprise a
zero energy sink which can however correlate both with the state and the
history of the system of interest, leading to loss of coherence without loss
of energy. This phenomenon plays a crucial role in the physics of such systems.

Our first model involves the interaction of
a quantum field with a simple harmonic oscillator, where the frequency
of the oscillator depends upon the state of a spin system. (Both the
oscillator and the spin system comprise the ``hidden system" in this
model.) The resonant
scattering of the field will thus depend on the state of the spin system,
leading to decoherence for that scattering process. By placing the
harmonic oscillator at some fixed point in space, the interaction and
loss of coherence will clearly be local in space, but the complete system
will fail to be translation invariant. Consequently, this model will
be one in which the field loses coherence without any violation of
causality or
energy conservation, but where momentum conservation fails.

Our second model is simply ordinary $\lambda \phi^4$ field theory
except that we now treat $\lambda$ as a random variable
 (which we may view as representing hidden degrees of freedom), with
probability distribution $R(\lambda)$.
In this case, loss of quantum coherence
will occur in scattering in such a way as to be entirely causal
and satisfy conservation of energy and momentum.

For our first model we consider a scalar field in one spatial dimension
which interacts with a
simple harmonic oscillator located at the origin. Radiation can excite the
oscillator, which then subsequently decays, reemitting its radiation. Such a
model clearly conserves energy, since all of
the energy absorbed by the oscillator
is reemitted. However, in this bare form, there is no loss of
coherence, since the field eventually
regains all of the coherence which is lost in the intermediate
states when one traces over the oscillator, as has been analysed by
Anglin,
Laflamme,  Zurek, and Pas \cite{anglin}. Indeed, this model is
a simplified form of a model in which a
lump of matter is heated by radiation, and then
cools down again. To create a genuine
loss of coherence, we couple the oscillator to an internal
system, which for simplicity, we take to be a spin system with
total spin $s$.
The spin system is assumed to have no free Hamiltonian, but gains an
energy only if the oscillator is excited. The total Hamiltonian is taken to be
\begin{equation}
H= {1\over 2} \int \left( (\pi(t,x) - h(x) q)^2
+(\partial_x\phi(t,x))^2 \right)dx
+{\omega\over 2}(p^2+q^2-{1\over 2})(1+\alpha(S_z+s)) +F(S_z)
\end{equation}
where the function $h(x)$ is sharply peaked around $x=0$, and will be treated
 as being proportional to a $\delta$ function. $S_z$ is the $z$ component
 of the spin operator for the spin $s$ system.
Note that $H$ commutes with $S_z$, so we may consider the
spectrum of $H$ separately in each sector of eigenstates
of $S_z$.
We choose the operator $F(S_z)$ so that the energies
of the lowest lying state in each of these sectors are identical.

Now, suppose that the spin system is in an eigenstate of the operator $S_z$
with eigenvalue $m$.
Then the energy levels of the oscillator are of magnitude
$\omega ((m+s)\alpha)$.
That oscillator will therefore absorb and reemit radiation
with frequencies around $\omega((m+s)\alpha)$. Thus, the scattered radiation
from that oscillator will depend on the state of the hidden
spin. This implies that if this spin system is not initially in an eigenstate
of $S_z$, correlations will develop between the spin system and the
quantum field. If we trace over the states of this spin,
the field will, in general evolve into a mixed state.

To see this explicitly, we note that
the solution to the equations of motion
for the above model (taking $h(x)=h\delta(x)$)
is given by
\begin{equation}
\phi(t,x)= \phi_{in}(t,x) +{1\over 2} h q(t-|x'|) dx'
\end{equation}
\begin{equation}
q(t,x) = \int^t e^{-\gamma (t-t')} {1\over \Omega_m}\sin(\Omega_m (t-t'))
h \dot\phi_{in}(t',0)dt'
\end{equation}
where $\phi_{in}$ is a free field operator, $\gamma= h^2/4$,
and $\Omega_m^2= (\omega(\alpha ( m+s))^2 -\gamma^2$.
Thus we have
\begin{equation}
\phi(t,x)=\phi_{in}(t,x)
+{\gamma\over \Omega_m} \int^t e^{-\gamma(t-t')}\sin(\Omega_m(t-t'))\dot
\phi_{in}(t',0) dt'
\end{equation}

Examining the scattering states, we find that the outgoing annihilation
operators of the field are related to the ingoing annihilation operators
by
\begin{equation}
a_{k,out}= a_{k,in}
 - i\sigma_{k,m} {(a_{ k,in}+a_{-k,in})}
\end{equation}
or reversing the flow,
\begin{equation}
a_{k,in}= a_{k,out}
+ i\sigma_{k,m} (a_{k,out}+a_{-k,out})
\end{equation}
where
\begin{equation}
\sigma_{k,m}= i{2\gamma|k| \over {-|k|^2+2i\gamma|k|+\Omega_m^2}}
\end{equation}
Thus, the outgoing state $|\psi_m>$ when the field initially
is in the state $\int \beta(k)a^\dagger_{k,in}|0>$,
the harmonic oscillator initially is in its ground state, and the
spin system initially is in the state $|m>$, is given by
 \begin{equation}
|\psi_m> = \int \beta_k (a_{k,out}^\dagger +\sigma^*_{k,m}
(a^\dagger_{k,out}+a^\dagger_{-k,out})) dk |0>
\end{equation}

Clearly, the complete model is causal (the field propagates away
from the oscillator
at the speed of light), energy conserving, and unitary.
However, when we start with the spin system in the state
$|\kappa>=\sum_m \kappa_m |m>$
the reduced density matrix of the field at very late times will be given
by
\begin{equation}
\rho= \sum_m |\kappa_m|^2 |\psi_m><\psi_m|
\end{equation}
To see at least an effect of the loss of coherence we can look at
\begin{equation}
Tr(\rho^2)= \sum_m\sum_{m'}|\kappa_m\kappa_{m'}|^2 |<\psi_m|\psi_{m'}>|^2
\end{equation}
the square of the density matrix, whose deviation from unity
gives a measure of the coherence lost in the scattering process.
We have
\begin{eqnarray}
<\psi_m|\psi_{m'}> &~&
\\
&=& < 0| \int \beta^*(k)\left(a_{out,k}
+\sigma^*_{|k|,m}(a_{out,k}+a_{out,-k})dk\right.
 \\
&&~~~~~~~~~~~~~~~~~~
\times \int \beta(k')\left(a^\dagger_{out,k}
 +\sigma_{|k|,m'}(a^\dagger_{out,k'}+a^\dagger_{out,-k'}\right)dk'|0>
\\
&=&\int |\beta(k)|^2 \left((1+\sigma^*_{|k|,m}
+\sigma_{|k|,m'} + \sigma^*_{|k|,m}\sigma_{|k|,m'})\right.
 \\
&&~~~~~~~~~~~~~~~~~ \left.
+\sigma^*_{|k|,m}\sigma_{|k|,m'}\right) dk
\end{eqnarray}
where we have assumed that $\beta(k)$ is non-zero only for $k>0$.
Thus we finally have
\begin{eqnarray}
Tr(\rho^2)&=& 1 + \sum_m 2|\kappa_m|^2 \int |\beta(k)|^2Re(\sigma_{|k|,m})dk
\\
&&~~~~~+2\sum_{m,m'}|\kappa_m|^2|\kappa_{m'}|^2
\int |\beta(k)|^2 \sigma^*_{|k|,m}\sigma_{|k|,m'}
dk
\end{eqnarray}
Since $<\psi_m|\psi_m> =1$, we also have that
\begin{equation}
2\int |\beta(k)|^2Re(\sigma_{|k|,m})dk =
-\int |\beta(k)|^2 \sigma^*_{|k|,m}\sigma_{|k|,m}
dk
\end{equation}
so that
\begin{equation}
Tr(\rho^2) = 1 - 2 \sum_{mm'} |\kappa_m|^2|\kappa_{m'}|^2 \int |\beta(k)|^2
\left(|\sigma_{|k|,m}|^2 -\sigma_{|k|,m}\sigma^*_{|k|,m'}\right) dk
\end{equation}
Thus, if the various $\sigma_{k,m}$ do not overlap, this becomes
\begin{equation}
Tr(\rho^2)=1- 2\sum_m (|\kappa_m|^2-|\kappa_m|^4) \int |\beta(k)|^2
\sigma^*_{|k|,m}\sigma_{|k|,m} dk
\end{equation}
If they do overlap, the sum will be increased from this expression. (If
they are independent of $m$, $Tr(\rho^2)$ is just unity, as would be expected.)
We can thus have a large reduction in coherence if the various $\kappa_m$
are sufficiently small.

For our second model, we start with ordinary $\lambda \phi^4$
field theory, and we let $U_\lambda (t_0,t)$ denote the unitary time
evolution operator of this theory. We now define a new dynamical
evolution law by choosing a time $t_0$ (which we may take to be
$t_0 = - \infty$) and setting
\begin{equation}
{\rho^A}_B (t) = \${^A}{_B}{_C}{^D}(t_0,t) {\rho^C}_D (t_0)
\label{evolph}
\end{equation}
where
\begin{equation}
\${^A}{_B}{_C}{^D}(t_0,t) =  \int R(\lambda) U^A_{\lambda C}(t_0,t)
{U^{\dagger D}_{\lambda}}_B(t_0,t) d\lambda
\label{plaw}
\end{equation}
and $R(\lambda)$ is an arbitrarily chosen probability distribution,
(so $R(\lambda)$ is non-negative and $\int R(\lambda) d\lambda = 1$).
Thus, the new dynamical evolution law corresponds to $\lambda \phi^4$
theory with $\lambda$ being a random variable. (Note that the dynamical
evolution law for the quantum field obtained in the above oscillator model
also is of the general form (\ref{plaw}),
and the non-Markovian character of both models is manifested
by the fact that  $\$(t_0,t) \neq \$(t_1,t) \$(t_0,t_1)$.)
It is easily seen that this
dynamical evolution law takes
initial pure states to mixed states. Indeed, for a pure initial state
$|\Psi_0>$ at time $t_0$,
the state at time $t$ is given by
\begin{equation}
\rho(t) = \int R(\lambda) |\Psi(t; \lambda)> <\Psi(t; \lambda)| d\lambda
\end{equation}
where $|\Psi(t; \lambda)> = U_\lambda (t_0,t) |\Psi_0>$. We have
\begin{equation}
Tr\rho^2 = \int\int R(\lambda)R(\lambda')
|<\Psi(t; \lambda)|\Psi(t; \lambda')>|^2
d\lambda d\lambda'
\end{equation}
which is not equal to unity unless the states $|\Psi(t; \lambda)>$
are independent of $\lambda$.

Since $\lambda \phi^4$ theory is causal for each value of $\lambda$, it
is manifest that the new dynamics is causal, i.e., no observers can use the
$\phi$-field to send signals faster than light. Furthermore, since energy
and momentum are conserved for each $\lambda$ and these observables
may be replaced by their free field ($\lambda = 0$) values after the
interactions have occurred, we see that energy and momentum (as well
as angular momentum) are exactly conserved in all scattering
experiments. The model also is Lorentz covariant. Thus, this model
explicitly demonstrates that loss of quantum coherence is not
incompatible with all of the above properties.

It should be noted that this model (as well as the previous oscillator
model) would have some unsatisfactory features
as a model of the black hole evaporation process.
Specifically, one would expect the results of an experiment involving
black hole formation and evaporation to be uncorrelated with the results
of similar experiments performed at other times or places. However, since
$\lambda$ is constant over spacetime, such correlations will occur in
this model, and information loss will not be ``repeatable'' \cite{gid}.
However, as we have discussed above, the
loss of quantum coherence in
processes involving a black hole may have many features which
cannot be modeled by a simple system where spacetime is treated
classically, so that coherence must be lost by
being transfered to some internal environment rather than
by falling into a
spacetime singularity.  Our purpose in this Appendix
was merely
to demonstrate that it is not necessary to violate causality or energy and
momentum conservation in order to have a loss of quantum coherence.


\begin{thebibliography}{99}
\bibitem{com1} It should be noted that the non-unitary character of the
Schrodinger picture evolution law eq.(\ref{eom}) is reflected in the
Heisenberg evolution law (\ref{heom}) by the fact that for two operators,
$A$ and $B$, we have, in general, $(AB)_H \neq A_H B_H$ (where the
subscript $H$ denotes the Heisenberg picture representative). In
particular, the Heisenberg evolution of a projection operator need not
remain a projection operator. The probabilities for the possible outcomes of
measuring an observable $A$ must be computed by Heisenberg
evolving the projection operators in the spectral family of $A$ rather
than by computing the spectral decomposition of $A_H$.
\end{thebibliography}
\end{document}